# Monte Carlo simulation of the experiment MAMBO I and possible correction of neutron lifetime result

A.P. Serebrov[1], A.K. Fomin

*Petersburg Nuclear Physics Institute, Russian Academy of Sciences, Gatchina, Leningrad District, 188300, Russia*

[1] *Corresponding author:* A.P. Serebrov
A.P. Serebrov
Petersburg Nuclear Physics Institute
Gatchina, Leningrad district
188300 Russia
Telephone: +7 81371 46001
Fax: +7 81371 30072
E-mail: serebrov@pnpi.spb.ru

**Abstract**

We are discussing the present situation with neutron lifetime measurements. There is a serious discrepancy between the previous experiments and the recent precise experiment [1]. The possible reason of the discrepancy can be connected with a quasi-elastic scattering of UCN on the surface of liquid fomblin which was used for most of the previous experiments. The Monte Carlo simulation of one of the previous experiments [2] shows that the result of this experiment [2] can be corrected and instead of the previous result 887.6 ± 3 s the new result 880.4 ± 3 s could be claimed.

*Keywords*: ultracold neutrons; neutron lifetime

**Introduction**

Recently the new experiment for the measurement of neutron lifetime [1] has been carried out with a high accuracy (0.8 s). The result of this experiment 878.5 ± 0.8 s is different from the world average value 885.7 ± 0.8 s presented in PDG 2006. This difference is considerable, 7.2 s or 6.5 standard deviations. The new experiment [1] with a gravitational trap and a low temperature fomblin oil coating of the trap walls has a few advantages in comparison with the previous experiments. First of all it is a low loss factor, $2 \cdot 10^{-6}$ per collision of UCN with trap walls. As a result the probability of losses was about 1% in comparison with the probability of neutron β-decay. Therefore the measurement of neutron lifetime was almost direct, the extrapolation from the best storage time to neutron lifetime was about 5 s only. In these conditions it is practically impossible to obtain a systematical error of about 7 s. The systematical error of the experimental result [1] was 0.3 s.

In connection with this situation we decided to analyze the previous experiments in more details. Most of the previous experiments with UCN storage have been carried out using fomblin oil coating at the room temperature in comparison with the experiment [1] which has been carried out at the temperature of ~120 K. The loss factor was about 30 times higher than in the experiment [1]. Very soon after these experiments [2-5] it was observed [6,7] that there is a quasi-elastic scattering of UCN in the reflection from the fomblin surface, which is rather extensive at the room temperature and can be suppressed at a low temperature [7]. This quasi-elastic scattering is happening due to surface waves of the liquid as it was proposed in the work [8]. The theoretical statements of this work have been checked experimentally in the laser experiments [8] and in the UCN experiment [7]. There is a reasonable agreement between theoretical calculations and experimental results devoted to the observation of low energy heating of UCN during the storage process in the trap with fomblin oil coating [7]. Therefore we decided to use the theory of quasi-elastic scattering process and realize Monte Carlo simulation of the first experiment [2] with the fomblin oil coating.

**Neutron lifetime experiment [2] scheme and results**

Below we reproduce a short description and results of the experiment [2]. The scheme of this experiment is shown in Fig. 1. The UCN storage volume is a rectangular box, with constant height =30 cm and width =40 cm but variable length $x<$ 55 cm. The side walls and the roof of the box are made of 5-mm float-glass plates. The oil spray head is mounted on the metal base plate and the assembly is immersed in a 1-mm-deep lake of oil. The movable rear wall, composed of two glass plates with a 1-mm oil-filled gap in between, has a 0.1-mm play with respect to the neighboring walls, except for the base plate where it dips into the oil. The surface of the rear wall was covered with 2-mm-deep, 2-mm-wide sinusoidal corrugations. For half the surface the wave crests were horizontal, and for the other half vertical. This arrangement transforms within a few seconds the forwardly directed incoming neutron flux into the isotropic distribution essential for the validity of the mean-free-path formula $\lambda = 4V/S$. The UCN inlet and outlet shutters situated 8 cm above floor level are sliding glass plates with 65-mm holes matching holes in the front wall (Fig. 1). In more details the description of experiment can be found in the work [2].

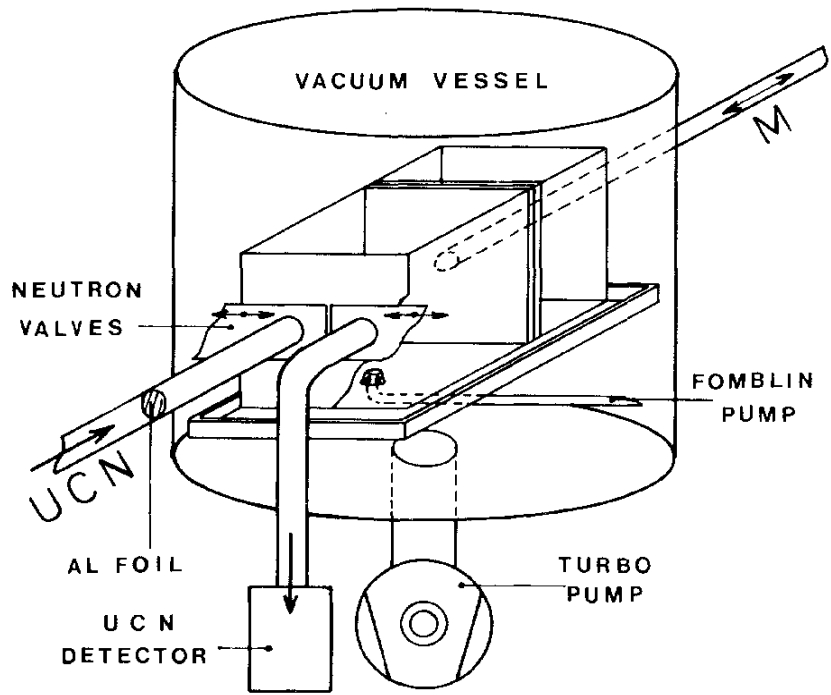

Fig. 1. Sketch of the apparatus MAMBO I.

The main idea of experiment [2] is to extrapolate UCN storage time to neutron lifetime changing the mean free path of UCN between wall collisions:

$$\tau_{st}^{-1} = \tau_n^{-1} + \mu(v)\nu(v) = \tau_n^{-1} + \mu(v)v/\lambda, \quad (1)$$

where $\tau_{st}^{-1}$ is probability of storage or inversed storage time, $\mu(v)$ is UCN loss factor per collision that depends from neutron velocity $v$, $\nu(v)$ is frequency of collisions, $\mu(v)\nu(v)$ is probability of UCN losses, $\lambda$ is mean free path, $\lambda = 4V/S$ for an isotropic and homogeneous particle population in a trap of volume $V$ and surface area $S$. Formula (1) shows that inverse storage time is linear function of inversed mean free path. The extrapolation of the inversed storage time to zero value of the inversed mean free path (or zero frequency of collisions) will give probability of neutron ß-decay.

Unfortunately, the correct extrapolation is impossible in case of wide UCN spectrum. UCN spectrum changes its form during the storage process due to dependence of $\mu(v)$ and $\nu(v)$ from neutron velocity. In work [2] it was proposed to fix number of collisions for different sizes of trap due to changing the holding time in trap by corresponding way. It helps to solve above mentioned problem. This scaling condition gives the following relations:

$$\frac{t_2(i)}{t_2(j)} = \frac{t_1(i)}{t_1(j)} = \frac{\lambda(i)}{\lambda(j)} = \frac{t_2(i) - t_1(i)}{t_2(j) - t_1(j)}, \quad (2)$$

where $t_1$ and $t_2$ are two different UCN holding times in the trap to determine $\tau_{st}^{-1}$, indexes $(i, j)$ corresponds to different volumes. Unfortunately even in case of the scaling conditions of measurements the extrapolation is violated because of gravitational field. This correction has to be calculated and included in final result for neutron lifetime. It was done in article [2].

Quasi-elastic scattering of UCN on the surface of liquid fomblin changes the UCN spectrum it violates the correctness of extrapolation also. The similar problem arises due to above barrier neutrons. Here we are going to estimate this additional correction which was not taken into account in article [2].

**Monte Carlo simulation of experiment [2]**

The MC simulation of the experiment [2] was done with a code which takes into account the effect of gravity and the effect of quasi-elastic scattering of UCN in the reflection from the fomblin coating. The dependence of the probability of quasi-elastic scattering from the initial and final energy of UCN was approximated in an analytical form using the theoretical dependences from the work [8]. The calculations were done on the computing clusters. The total time of calculations is about several months. The UCN storage volume was a rectangular box with a variable length $x$. The dimensions of the box were the same like in the experiment. In the following all detailed results are given for the reference volume with $x =55$ cm unless stated otherwise. Neutrons in the trap have specular (50%) and diffuse (50%) reflections from the walls to simulate the corrugated surface of the rear wall in the experiment. The time intervals of UCN storage have been chosen the same as in the experiment. Fig. 2 shows the simulated data in comparison with the experimental ones. The same way as in the experiment the storage time (or inversed storage time) was extrapolated to neutron lifetime (or probability of neutron β-decay). In our simulations neutron lifetime was exactly fixed. Therefore we can calculate the correction to the extrapolated neutron lifetime.

As the first step of our calculations or as benchmark test of our model we decided to reproduce the main experimental dependence of extrapolated neutron lifetime from storage time intervals (Fig. 3 from article [2]). This benchmark test is shown in Fig. 3 of our article in comparison with experimental result of article [2]. We can reproduce the main dependence and also experimental effect due to changes of mirror reflectivity of trap surface.

The next question was how result of calculation depends from initial UCN spectrum. It was important question because form of initial UCN spectrum was known in general only. Fig. 4 demonstrates that result of calculations (particularly for the most important points with long holding time) do not practically depend from the form of initial UCN spectrum.

Then we started to study effects of quasi-elastic scattering and above barrier neutrons. The gravitational correction was calculated also.

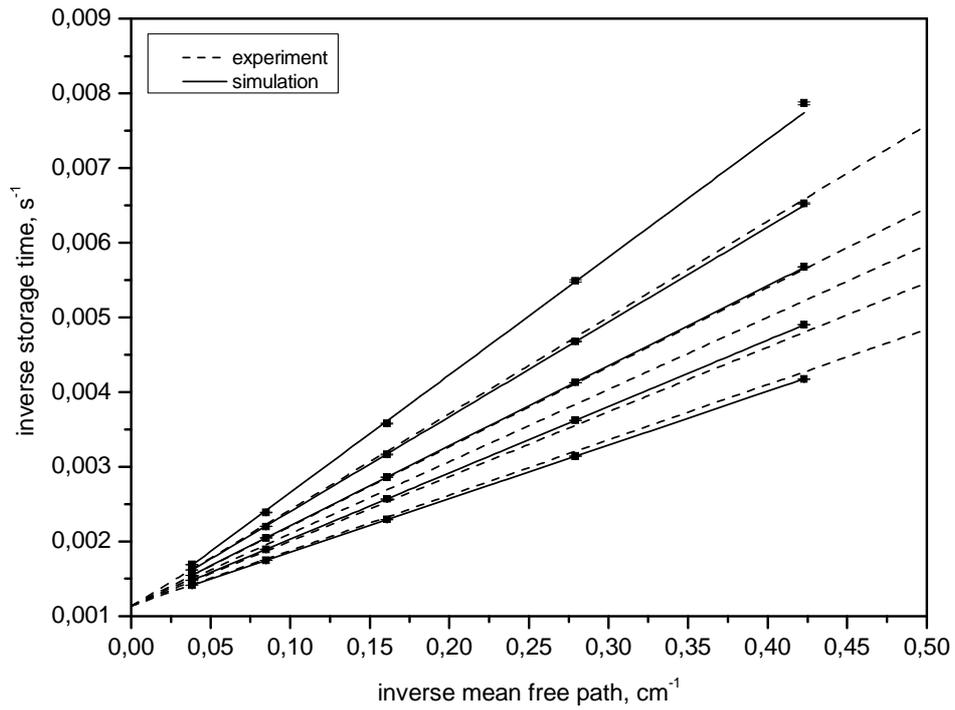

Fig. 2. Measured inverse bottle lifetimes as a function of the bottle inverse mean free path and for different storage intervals (dotted lines). Simulated data (solid lines) in comparison with experimental ones.

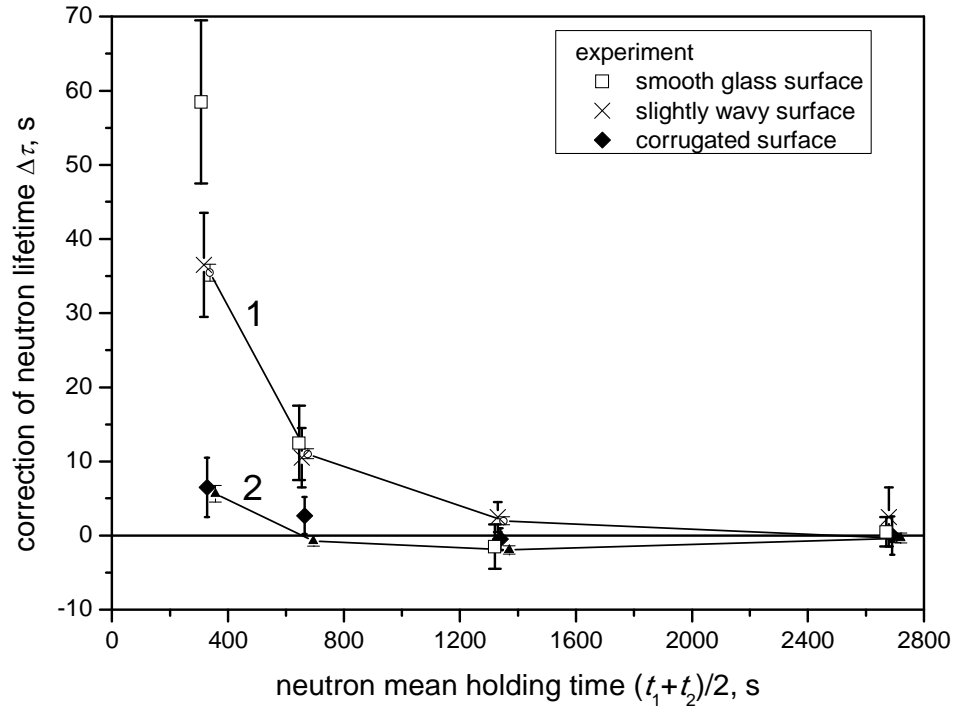

Fig. 3. Dependence of the experimental uncorrected neutron lifetime on the storage time intervals for different bottle surface structures in comparison with results of the simulations with different probability of diffuse reflections from the walls. (1) surface with diffuse reflections 1%, (2) surface with diffuse reflections 50%.

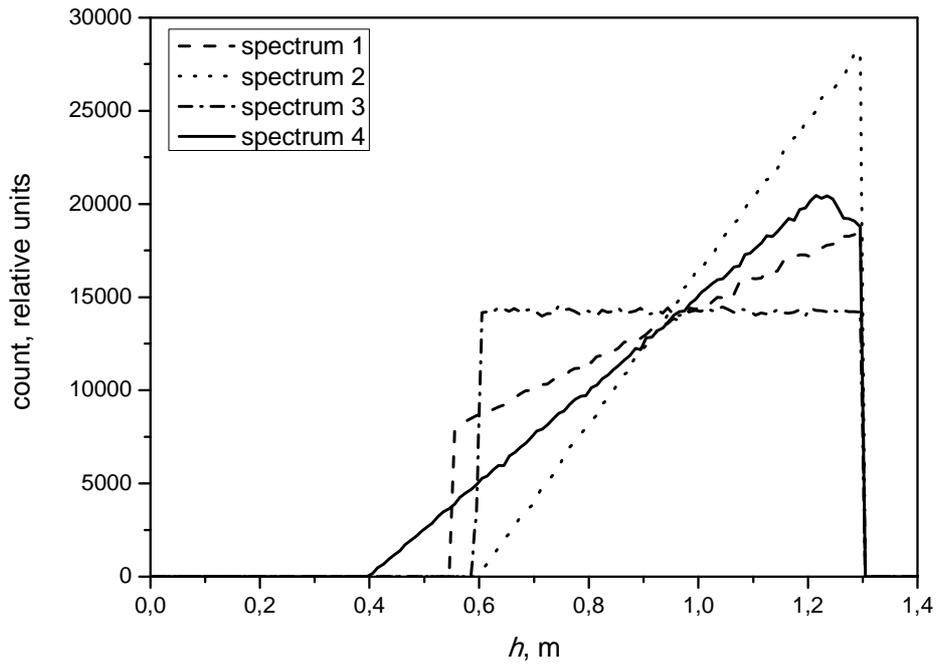

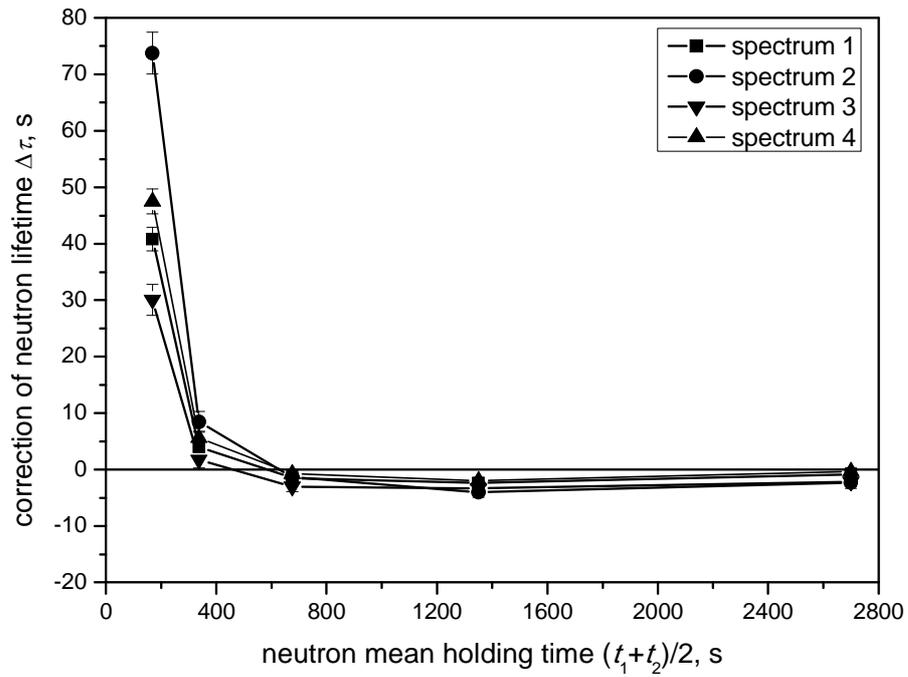

Fig. 4. Results of the simulations with different initial UCN spectrums in the trap.

In this experiment there are two main problems which have to be considered. The first is that the initial UCN spectrum contains above barrier neutrons which can be stored for a long time particularly if the energy is near the critical one. It can cause a systematic error. The second one is the quasi-elastic scattering which changes the form of the spectrum during the storage process. It can destroy the condition of scaling declared in the work [2]. The effect of the spectral changes during the storage process is shown in Fig. 5. One can see that quasi-elastic scattering changes the form of UCN spectrum considerably.

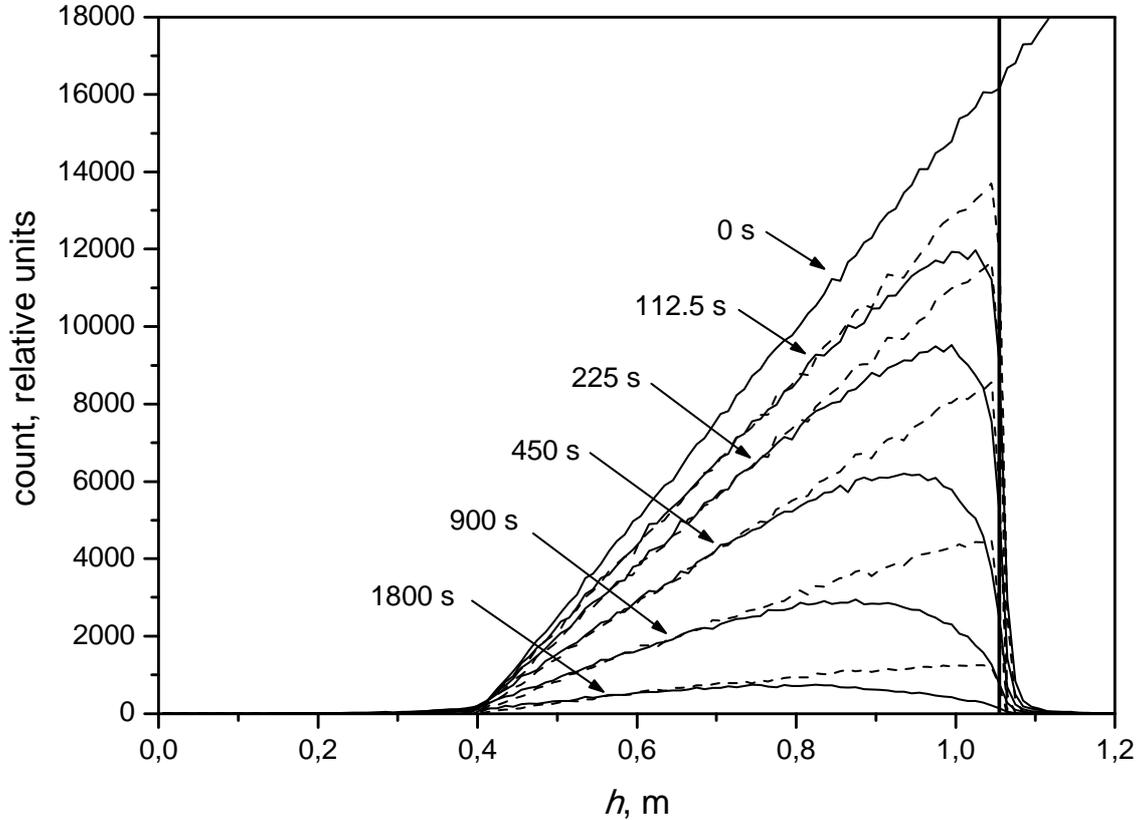

Fig. 5. UCN spectra in the trap after different storage intervals without taking into account quasi-elastic scattering (dotted lines) and with taking into account quasi-elastic scattering (solid lines).

The results of extrapolations to neutron lifetime are shown in Fig. 6 for different conditions of measurements: with taking into account above barrier neutrons and without above barrier neutrons, with taking into account quasi-elastic scattering and without quasi-elastic scattering.

In case when the above barrier neutrons are absent and there is no quasi-elastic scattering (curve 1) we can calculate the gravitational correction only. The bigger volumes have relatively larger area of bottom and more collisions with higher energy due to gravity. It gives lower value of extrapolated neutron lifetime. The gravitational correction is practically independent from the UCN holding time in the trap. The extrapolated neutron lifetime is found lower than neutron lifetime by $7.5 \pm 0.3$ s. This result is similar to the gravitational correction introduced in the work [2].

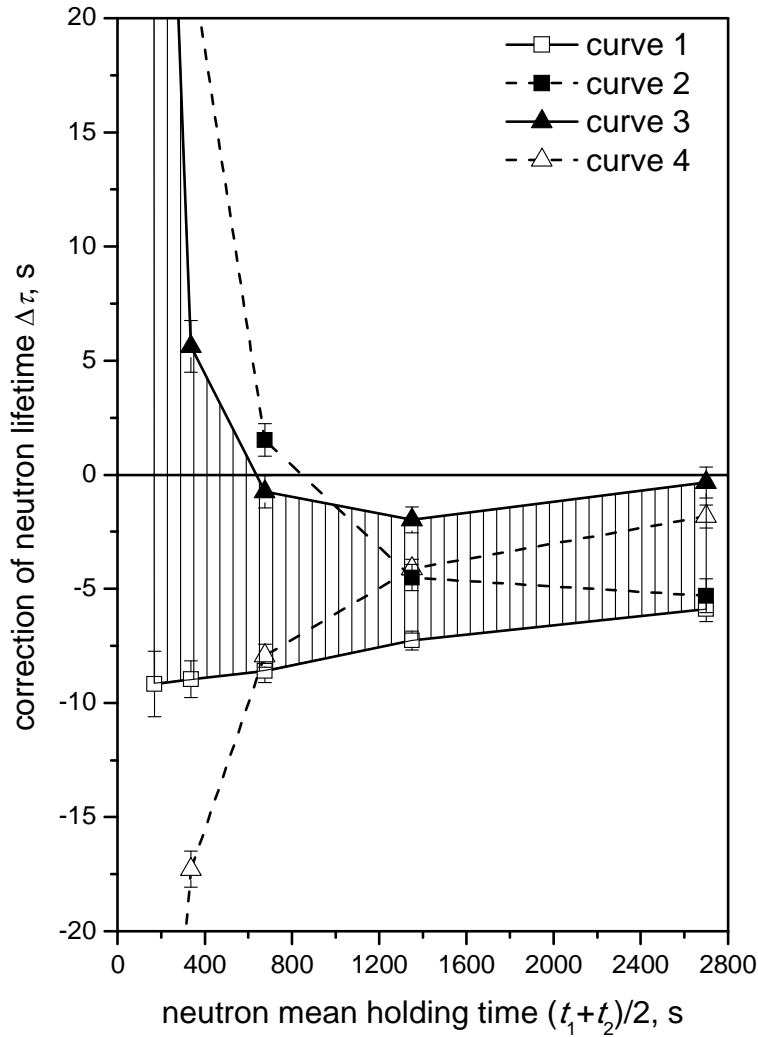

Fig. 6. Results of MC simulations of the extrapolated neutron lifetime for different storage intervals. (1) without quasi-elastic scattering and without above barrier neutrons, (2) without quasi-elastic scattering and with above barrier neutrons, (3) with quasi-elastic scattering and with above barrier neutrons, (4) with quasi-elastic scattering and without above barrier neutrons. The difference between curves 1 and 3 is correction due to above barrier neutrons and quasi-elastic scattering which was not taken into account in work [2].

The next simulation was done with a full UCN spectrum including above barrier neutrons but without quasi-elastic scattering (curve 2). One can see that for the short holding time the extrapolated neutron lifetime is much higher in comparison with the previous case (curve 1), but for the long storage time the extrapolated neutron lifetime became rather close to the curve 1 for the gravitational correction. It should be mentioned that the contribution of results with a short holding time in the final result is very small because of poor statistical accuracy of these measurements in the experiment.

The points with a holding time of (900-1800) s and (1800-3600) s bring the main contribution.

The next simulation was done with taking into account quasi-elastic scattering and above barrier neutrons (curve 3). The difference between curves 1 and 3 is the total effect due to above barrier neutrons and quasi-elastic scattering. These effects were not taken into account in the work [2]. Table 1 is the table from the work [2] with our additional corrections due to both effects. The total correction is $-7.2 \pm 1.6$ s and corrected result for neutron lifetime is $880.4 \pm 3$ s. This result is not in contradiction with the result of the work [1] $878.5 \pm 0.8$ s.

Table 1. Results of $\tau_n$ for different storage intervals: $\tau_n$ is corrected result for neutron lifetime from work [2], $\Delta\tau$ is correction due to above barrier neutrons and quasi-elastic scattering calculated in this work, $\tau'_n$ is result for neutron lifetime after taking into account correction $\Delta\tau$ from this work.

| storage interval, s | $\tau_n$, s | $\Delta\tau$, s | $\tau'_n$, s |
|---|---|---|---|
| 112.5-225 | 891(10) | -56.68 (2.63) | 834.32 (10.34) |
| 225-450 | 888.5(4) | -14.58 (1.39) | 873.92 (4.23) |
| 450-900 | 889.2(2.5) | -7.84 (0.87) | 881.36 (2.65) |
| 900-1800 | 887.0(1.5) | -5.29 (0.70) | 881.71 (1.65) |
| 1800-3600 | 887.1(2.6) | -5.54 (0.87) | 881.56 (2.74) |
|  | $\tau_n$ =887.6(1.1) s |  | $\tau'_n$ =880.4(1.2) s |

Finishing the article we have to say that official correction of the result of experiment [2] is matter of authors. We have no rights to do this. Our task is to show possible correction which was unknown in course of experiment [2] realization.

In conclusion we would like to give a high regard to the main initiator of the experiment [2], Walter Mampe, who made a very significant contribution to the development of UCN experiments at ILL, and who succeeded in uniting the physicists from different countries in order to carry out these tasks. We are very thankful to Mike Pendlebury for useful discussions in course of this work. The given investigation has been supported by RFBR grant 07-02-00859. The calculations were done at computing clusters: PNPI ITAD cluster, PNPI PC Farm.